\documentclass[runningheads]{llncs}
\usepackage{graphicx}
\usepackage{booktabs,microtype,hyperref,xltabular,float,array}
\usepackage{makecell}
\hypersetup{colorlinks,
      linkcolor=blue,
      citecolor=blue,
      urlcolor=blue,
      filecolor=blue}
\usepackage[square, comma, sort&compress, numbers]{natbib}

\usepackage[usenames,dvipsnames]{xcolor}
\usepackage[color=RoyalBlue,textcolor=white]{todonotes}
\usepackage{csquotes}
\usepackage{cleveref}

\usepackage{tikz}
\usepackage{enumitem}
\usepackage{soul}

\def\defineCMYKcolor(#1,#2,#3,#4)#5{%
    \pgfmathsetmacro{\myc}{#1/100}%
    \pgfmathsetmacro{\mym}{#2/100}%
    \pgfmathsetmacro{\myy}{#3/100}%
    \pgfmathsetmacro{\myk}{#4/100}%
    \definecolor{#5}{cmyk}{\myc,\mym,\myy,\myk}%
}

\defineCMYKcolor(44, 76, 0, 29){purplef}
\definecolor{magentaf}{HTML}{f8cce0}
\defineCMYKcolor(0, 100, 55, 14){magentaf2}
\definecolor{greenf}{HTML}{e9f6dc}
\defineCMYKcolor(92, 0, 18, 35){greenf2}
\definecolor{orangef}{HTML}{fcebd4}
\defineCMYKcolor(0, 33, 80, 8){orangef2}
\definecolor{bluef}{HTML}{d9dbf0}
\defineCMYKcolor(61, 55, 0, 34){bluef2}
\definecolor{brownf}{HTML}{eddfdf}
\defineCMYKcolor(0, 40, 42, 37){brownf2}
\defineCMYKcolor(0, 4, 33, 4){vanillaf}
\definecolor{melonf}{HTML}{ffbbad}
\defineCMYKcolor(0, 27, 32, 0){melonf2}
\definecolor{plumf}{HTML}{eaaede}
\defineCMYKcolor(0, 26, 5, 8){plumf2}
\definecolor{skybluef}{HTML}{99e1e6}
\defineCMYKcolor(33, 2, 0, 10){skybluef2}
\definecolor{objective}{HTML}{7e68b7}
\definecolor{subjective}{HTML}{7fb966}
\definecolor{experience}{HTML}{f88920}
\definecolor{interaction}{HTML}{1f77b4}
\definecolor{explanation}{HTML}{da0063}
\definecolor{situational}{HTML}{0CA789}
\definecolor{personal}{HTML}{CA6855}

\newcommand{\highlight}[2][]{\colorbox{#1!25}{\textit{#2}}}

\newcommand{\anon}[1]{\texttt{anonymized}}
\begin{document}
\title{Towards a Comprehensive Human-Centred Evaluation Framework for Explainable AI}
\titlerunning{A Human-Centred Evaluation Framework for Explainable AI}
%
\author{Ivania Donoso-Guzmán\inst{1,2}\orcidID{0000-0002-2427-9128} \and
Jeroen Ooge\inst{1}\orcidID{0000-0001-9820-7656} \and
Denis Parra \inst{2}\orcidID{0000-0001-9878-8761}\and 
Katrien Verbert\inst{1}\orcidID{0000-0001-6699-7710} 
}
\authorrunning{Ivania Donoso-Guzmán, Jeroen Ooge, et al.}
%
\institute{KU Leuven, Department of Computer Science \and
Pontificia Universidad Católica de Chile}
\maketitle              
\begin{abstract}
While research on explainable AI (XAI) is booming and explanation techniques have proven promising in many application domains, standardised human-centred evaluation procedures are still missing. In addition, current evaluation procedures do not assess XAI methods holistically in the sense that they do not treat explanations' effects on humans as a complex user experience. To tackle this challenge, we propose to adapt the User-Centric Evaluation Framework used in recommender systems: we integrate explanation aspects, summarise explanation properties, indicate relations between them, and categorise metrics that measure these properties. With this comprehensive evaluation framework, we hope to contribute to the human-centred standardisation of XAI evaluation.
\end{abstract}

\section{Introduction}
\textit{Explainable AI} (XAI) is advancing fast: between 2017 and 2021 alone, the number of XAI papers increased eight-fold \cite{Nauta2022FromAI} and researchers have proposed XAI methods for virtually all existing media types and families of AI models.
However, it is still unclear to what extent explanations are effective in practice \cite{Markus2020TheStrategies} because full-fledged standardised evaluation procedures are missing. This is partly due to lacking consensus on which explanation properties should be assessed and which measurements should be used 
\cite{Miller2019, Nauta2022FromAI, Vilone2021NotionsIntelligence, Carvalho2019MachineMetrics, Markus2020TheStrategies}. 

To better assess XAI methods, researchers have tried to disentangle explanation's characteristics into simple measurable properties such as completeness \cite{Nauta2022FromAI, Vilone2021NotionsIntelligence, Beckh2022ALearning}, novelty \cite{Sokol2020ExplainabilityApproaches, Liao2022ConnectingAI, Lofstrom2022AMethods, Carvalho2019MachineMetrics}, and interactivity \cite{Nauta2022FromAI, Hsiao2021RoadmapXAI, Vilone2021NotionsIntelligence}. However, there is little evidence on how these properties relate to explanations being appropriate in real scenarios \cite{Liao2021Human-CenteredExperiences}. In addition, while many researchers stress the importance of context, we are unaware of XAI evaluation methods that treat explanations' effects on humans as a \textbf{complex user experience} involving factors such as user perception and system interaction. 

To evaluate explanations holistically, we are working towards a human-centred evaluation framework for XAI, which extends pioneering work on developing and evaluating user experience \cite{Knijnenburg2015EvaluatingExperiments} and explanations \cite{Tintarev2015ExplainingEvaluation} for recommender systems. We categorise explanation properties according to this framework and indicate their relations reported in the literature. Additionally, we present the \textit{explanation elements} that help to classify metrics to simplify the choice of measurements. This adapted user-centric framework will allow researchers and practitioners to evaluate explanations of AI-based systems and potentially increase deployment of such systems in their respective domains \cite{Mohseni2021ASystems, Markus2020TheStrategies}.

The contributions of this paper are three-fold: first, we present an extensive analysis of existing definitions of explanation properties and methods, as well as their interrelationships. Our analysis aligns different properties and methods as defined by different research communities. Second, based on this analysis, we define a human-centred evaluation framework for XAI that presents an integrative approach and combines user-centric evaluation and functional metrics. Third, we present an example of the use of this framework.

\section{Background and Related work}

\subsection{Human-Centred Explainable AI}

The XAI area of research has been led mostly by the AI community, even though it is a multidisciplinary area of research. For this reason, XAI methods have been criticised for being developed with the AI researchers' intuition of what constitutes a good explanation \cite{Miller2019}. In particular, the design and evaluation of XAI methods are often conducted without considering the final users' needs and their cognitive processes \cite{Liao2021Human-CenteredExperiences}.

More recently, the HCI community started proposing ideas for tackling the XAI design, considering how the users reason about explanations: \citet{Wang2019DesigningAI} proposed a framework to design explanations based on how humans reason; \citet{Chen2022MachineUnderstanding} characterised how explanations affect human understanding of task decision boundary, model decision boundary and model error; Most recently, \citet{Chen2023UnderstandingExplanations} conducted a study to investigate the decision-making process users follow when faced with AI predictions and their explanations.

Another line of work has been understanding the wants and needs of different shareholders and ensuring they are considered in the design. \citet{Mohseni2021ASystems} categorised the goals of target user groups and developed design guidelines to iteratively design and evaluate Explainable AI systems; \citet{Suresh2021BeyondNeeds} proposed a framework to characterise users with two multidimensional criteria: knowledge and interpretability needs, that together help to understand the system's users; \citet{Langer2021WhatResearch} review the main types of users of XAI systems and their wants and needs, to propose a model for designing XAI systems according to these desiderata; \citet{Liao2021Question-DrivenExperiences} proposed a question-driven design process to fulfil the Explainable AI user's needs; \citet{Rong2022TowardsExplanations} analysed human-based XAI evaluations and provided guidelines for conducting user studies in the area.

Overall, these studies have emphasised the importance of  users' characteristics and the tasks they perform during the design phase of XAI experiences. Although it has been stated as an important aspect of the final adoption of XAI systems \cite{Rong2022TowardsExplanations, Mohseni2021ASystems}, to the best of our knowledge, evaluation procedures that capture the complexity of the human-AI interaction have not yet been proposed. We contribute by adapting a widely accepted procedure in recommender systems to evaluate explanations generated by XAI methods holistically.

\subsection{Evaluating Explanations}

Even though AI/ML models have standard evaluation metrics, there is still no consensus on the strategy to evaluate XAI methods. \citet{Doshi-Velez2017TowardsLearning} proposed the first standardisation of XAI evaluation. According to their work, the evaluation could be performed in three levels: application-grounded, with real tasks and users; human-grounded, with real users and proxy tasks; and functionality-grounded, with proxy tasks and no users. Currently, application or human-grounded approaches have been criticized for their lack of rigour \cite{Johs2020QualitativeScience, Johs2022ExplainableInvestigation}, and for using proxy tasks \cite{Bucinca2020ProxySystems}. 

To conduct functionally-grounded evaluations, i.e. proxy tasks and no users, some studies have focused on grouping concepts and defining properties 
\cite{Markus2020TheStrategies, Vilone2021NotionsIntelligence, Carvalho2019MachineMetrics, Beckh2022ALearning} and their corresponding metrics \cite{Nauta2022FromAI}. These works aggregate existing literature that defines properties or presents metrics to assess them. The proposed properties try to measure the quality of the explanations without context so that they can be used in functionality-grounded evaluation. Similarly, \citet{Hoffman2018MetricsProspects} proposed to evaluate explanations using the `goodness criteria' that assess the explanation quality without context. Most recently, \citet{Agarwal2022OpenXAI:Explanations} presented a framework to benchmark different XAI methods using automatic metrics. Still, it is limited to particular methods and only works with specific datasets created for the benchmark.

Little work has been conducted to present the connections between these properties. Most papers state that trade-offs exist \cite{Liao2022ConnectingAI, Nauta2022FromAI, Carvalho2019MachineMetrics, Markus2020TheStrategies}, but they have not quantified them. To the best of our knowledge, only the study by \citet{Balog2020MeasuringQuality} uncovered conflicting relationships between some of the proposed properties, but they did not evaluate XAI-generated explanations.

Given the number of properties to evaluate, selecting the aspects to consider in the evaluation is becoming an important topic. According to \citet{Liao2022ConnectingAI}, this selection depends on the tasks the system has to support because the user accomplishment of these tasks determines the overall system's success. \citet{Knijnenburg2015EvaluatingExperiments} indicate the selection is made according to theoretical models, i.e., it results from previous studies or from the hypothesis that is tested. Recently, \citet{Liao2022ConnectingAI} presented a study that connects tasks with evaluation criteria to provide general guidelines for the field. In this study, experts and end-users selected the most appropriate properties to evaluate diverse XAI tasks. They found that XAI tasks obtained different property rankings regardless of the application domain (loan application, medical diagnosis, among others). 

Our work builds upon these previous studies by proposing a unified framework that integrates previously proposed definitions and measurements by making the relations between them explicit and grounded in previous work. Additionally, we analysed measurement procedures and classified them by which explanation element they measure according to Miller's definition of explanation \cite{Miller2019}, which declares that explanations are composed of a cognitive process, a product and a social process. This new criteria to classify measurements provides researchers and practitioners with a new understanding of how to measure  properties of explanations.

\subsection{User Centric Evaluation of Recommender systems}

\begin{figure}
\centering
  \includegraphics[width=.75\linewidth]{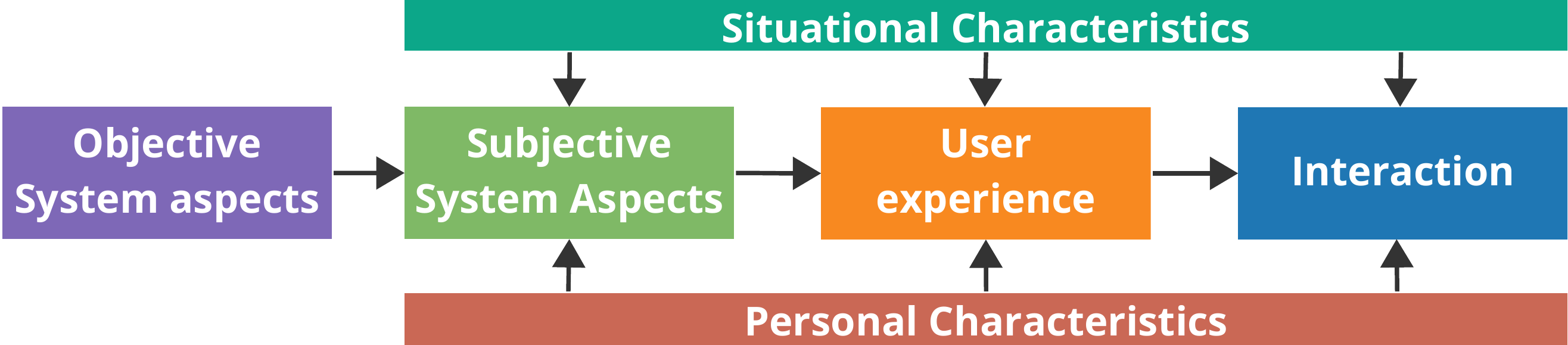}
  \caption[]{The User-Centric Evaluation Framework by Knijnenburg et al. \cite{Knijnenburg2015EvaluatingExperiments}. Each box represents a \textit{conceptual component} that groups related \textit{constructs}. 
  }
  \label{fig:original_framework}
  \vspace{-0.5cm}
\end{figure}

The User Centric Evaluation Framework for recommender systems in \Cref{fig:original_framework} was proposed by Knijnenburg et al. \cite{Knijnenburg2015EvaluatingExperiments} to explain how users experience the interaction with a recommendation system and to predict how users behave under similar circumstances. 
This framework has six \textit{conceptual components} encompassing different \textit{constructs} that can be measured during a user study. For example, the conceptual component \highlight[subjective]{\textit{Subjective system aspects}} groups constructs such as \textit{Perceived recommendation quality} or \textit{Interaction adequacy}, while \highlight[experience]{\textit{User experience}} contains \textit{Choice difficulty} and \textit{Choice satisfaction} among others. The constructs and the causal relations between them found with Structural Equation Modelling (SEM) \cite{Kline2023PrinciplesModeling} help explain how different aspects of the experience affect each other and influence the outcomes. 

This evaluation framework has been used and appreciated in recommender systems because of its capacity to provide relations between different user experience aspects. By capturing the causal relations between different measurements, researchers can not only report and compare these measurements but also explain why differences do or do not occur. This provides a better understanding of what makes a system more adapted to the users and, ultimately, predicts whether it will be successful and why.

In this work, we expand this successful framework for XAI evaluation. We believe our comprehensive work sheds light on which explanation aspects are more important and relevant to users and their circumstances. Furthermore, since the framework provides causal relations between different properties, we believe it can provide better guidelines for XAI design. 

\section{Methods}
To adapt the user-centric evaluation framework by \citet{Knijnenburg2015EvaluatingExperiments}, we analysed current literature on the topic with a grounded theory approach. This section describes how we collected papers and categorised them along two axes (conceptual components and explanation elements), to build the foundation for our XAI framework.

\subsection{Paper Collection}

Finding relevant literature on XAI evaluation requires searching several research disciplines. 
Evaluation, in particular, has been published in several types of venues (workshops, posters, surveys), presenting concrete methods and execution procedures but also proposals and blue-sky ideas. To include as much relevant literature as possible, we consulted Google Scholar with this query:

\begin{verbatim}
intitle:properties OR intitle:evaluation OR intitle:metrics 
OR intitle:property OR intitle:metric 
("explainable" OR "interpretable") 
("artificial intelligence" OR "machine learning" OR XAI OR AI)
\end{verbatim}

The search was conducted at the end of October 2022 and was limited to the years 2017 and onwards because \citet{Doshi-Velez2017TowardsLearning} then proposed one of the first XAI evaluation procedures. This query returned approximately 5970 results. As a first step, only the titles were reviewed to check whether the result was related to AI or XAI. We checked all result pages until the first page where no papers related to XAI or AI appeared. This occurred on page 25, similar to the results of \citet{Vilone2021NotionsIntelligence}. This first screening yielded 80 research works.

These works were analysed by looking at the abstract and, in doubt, at the full paper. The aim of this second screening was to remove duplicate works and keep only works that describe properties, relations between them and measurements. The exclusion criteria were the following:
\begin{itemize}
\item The research did not use or propose properties or measurements for XAI explanations.
\item The study considered only non-XAI-generated explanations.
\item The research compared different XAI methods using different metrics, but said metrics were not grounded on explanation quality aspects.
\item The evaluation of the explanations was performed with a ground truth explanation.
\item The search result was a master's or PhD thesis, and one or more papers were already published based on the same research, making it redundant.
\end{itemize}

After this screening process, only 19 results were kept. From their references, other related papers were found. We also included \cite{Tintarev2015ExplainingEvaluation} because it is a comprehensive review of the evaluation of explanations in the context of recommender systems. The final number of papers included was 29.

\subsection{Classification Axis 1: Conceptual Components}\label{sec:method:conceptual_components}

A Grounded Theory \cite{Charmaz2014ConstructingAnalysis} approach was followed to analyse the collected works in three steps: Initial Coding, aimed at finding quotes that related to properties of explanation; Focused Coding, which consisted of labelling the passages according to a set of concepts; and finally Axial Coding, which connects and groups the different concepts. 

The Initial Coding step was conducted in-vivo. Definitions of explanation properties, definitions of metrics to measure aspects of explanations, and relations between properties were searched for. Some of the papers had definitions of properties based on multiple previous works. In those cases, we kept the summarised definition and did not look for primary sources. In contrast, if the definition made in the survey paper did not fully explain metrics, we added the primary source to the group of papers.

The Focused Coding Step consisted of labelling the different definitions with the most appropriate concept, independently of the name the authors had coined. This iterative process aimed to group the definitions that point to the same desiderata of an explanation while avoiding overlapping concepts. The definition of each property was created at this step.  In addition, passages that described a procedure to measure the property were marked as such. The procedure to analyse those quotations is described in \Cref{sec:method:explanation_elements}.

The Axial Coding phase was conducted by first collecting the relations that were described in the selected papers. After these relations were captured, new relations that emerged from the definitions were investigated and added to the model. Additionally, relations were added based on evidence of other papers the researchers were aware of.

\begin{figure}[h]
\centering
\includegraphics[width=\linewidth]{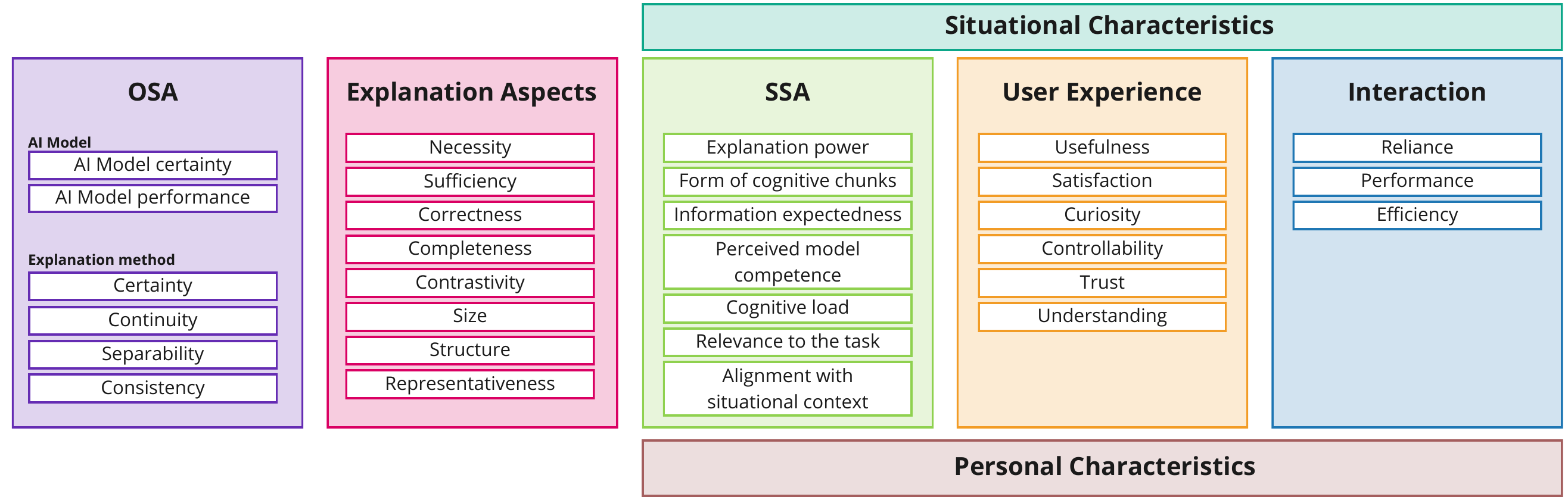}
  \caption{The User-Centric Evaluation Framework by Knijnenburg et al. \cite{Knijnenburg2015EvaluatingExperiments} extended with a new conceptual component: \textit{Explanation Aspects}. Each conceptual component displays its properties. The box of Objective system aspects (OSA) marks the properties that apply to the AI model and the ones that apply to the XAI method.}
\label{fig:framework_overview}
\vspace{-0.5cm}
\end{figure}

Finally, each of the found properties was matched to a conceptual component as defined in Knijnenburg's framework \cite{Knijnenburg2015EvaluatingExperiments}. Our analysis yielded very few and general properties for the situational and personal characteristics components, so it was decided to leave those properties out of the current analysis. During this phase, it was noted that some properties belonged to a new category that captured the abstract quality of the explanation. This idea aligns with the nature of XAI methods: the original framework was made for recommender systems, i.e., an AI model that selects objects, but XAI methods \textbf{generate} an object. To evaluate the quality of generated objects, it was decided to add the conceptual component \textit{Explanation Aspects} (see \Cref{fig:framework_overview}), which groups properties that evaluate the explanation quality.  

\subsection{Classification Axis 2: Explanation Elements}
\label{sec:method:explanation_elements}
\begin{figure}[h]
  \centering
  \includegraphics[width=\linewidth]{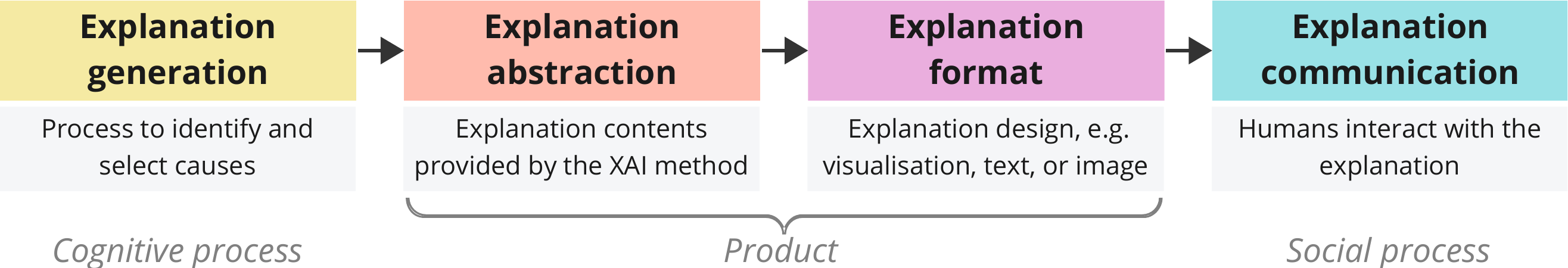}

  \caption[]{The four elements of explanations. We use the same ideas as {\cite{Miller2019}} but changed the names of the elements.Additionally, we further divide the explanation product into \textit{abstraction} and \textit{format}.}
  \label{fig:elements}
\end{figure}

Previous analysis of properties had classified measurement and metrics depending on their user dependency \cite{Beckh2022ALearning}, the nature of the procedure (objective, subjective)\cite{Hsiao2021RoadmapXAI, Coroama2022EvaluationXAI} or according to umbrella properties \cite{Moreira2022BenchmarkingBox, Nauta2022FromAI}. However, during the analysis of the conceptual components and the properties of explanations, it was found that similar properties are often named differently because of the ways in which they are measured. For example, \citet{Carvalho2019MachineMetrics} defined two similar concepts that were applied in two types of evaluation. They used the name \textit{Representativeness} for the evaluation without users and the concept \textit{General and probable} for evaluation with user studies, even though both refer to the number of instances that can be explained with the same causes. 
We argue this inconsistency occurs because explanations are made of different elements. Miller \cite{Miller2019} states that \textit{explanations} are both processes and products: the \textit{Cognitive process} selects a subset of the causes; the \textit{Product} is the resulting outcome; and the \textit{Social process} consists of transferring the knowledge from explainer to explainee. 

With these ideas in mind, a focused coding was conducted only of the passages marked as describing a procedure to measure a property. Each passage was labelled as \textit{generation}, \textit{product} or \textit{communication}. It was found that many metrics that were labelled \textit{product} were very format dependent: for example, BLEU (BiLingual Evaluation Understudy)\cite{Clinciu2021AExplanations}, which evaluates machine-translation quality, cannot be applied to visual-based explanations, but Covariate Homogeneity \cite{Nauta2022FromAI} could be applied to both text and visual-based explanations. For this reason, the metrics under the \textit{product} label were further categorised between \textit{abstraction} and \textit{format}. \Cref{fig:elements} displays the new definitions and the relation to Miller's definitions. 

This categorisation allows classifying measurement procedures under three criteria: property they measure, element of explanation and type of procedure (questionnaire, metrics, etc). Different measurements can be applied to evaluate the properties along the four explanation elements. Some properties can only be assessed by measuring one element, while others can be measured in more than one. These new criteria are explained and justified in \Cref{sec:framework_measurements}.

\section{A User-Centric Evaluation Framework for XAI}

This section presents an adapted version of the \textit{User-Centric Evaluation Framework}. To describe it, we use the following terminology: \textbf{conceptual components} group \textbf{explanation properties}, which in turn can be measured with \textbf{measurements}. While each measurement applies to only one \textbf{explanation element}, a single property can be measured by several measurements.

This section is organised as follows: in \Cref{sec:framework_properties}, the choice of properties for each conceptual component is justified and explained, and the properties are defined; then, in \Cref{sec:framework_relations}, the connections between properties are presented; finally in \Cref{sec:framework_measurements} the classification criteria for measurements is presented and justified, as well as the existing measurements for each property.

\subsection{Explanation Properties}\label{sec:framework_properties}

\subsubsection*{Objective system aspects.}
Objective systems aspects (OSAs) are `the aspects of the system that are currently being evaluated' \cite{Knijnenburg2015EvaluatingExperiments}. It was found from the analysis that characteristics from the particular instance of the XAI method and AI model can affect  the explanation. For instance, the AI model performance will affect the level of Trust users can achieve. Making these characteristics explicit in the framework can help to understand the specific aspects of the XAI method and AI model that affect the user experience.

The analysis yielded six properties: AI model performance, AI model certainty, Certainty, Continuity, Separability and Consistency. The first two properties measure the AI model, and the last four are applied to the XAI method. Continuity was described in several works as the desired `smoothness' of the XAI function. In the beginning, Separability and Continuity were one concept, but it was noted that providing similar explanations to similar instances does not guarantee that different instances will get different explanations. Consistency evaluates the randomness of the XAI method: if different runs of the XAI method algorithm return different functions, the model will be highly inconsistent. 

AI model certainty and XAI method certainty were complicated properties. 
Uncertainty quantification is a very active field of research within AI, and several approximation methods have been proposed. However, the problem is still being investigated due to its high computational cost \cite{Abdar2021AChallenges}. Papers' definitions for these concepts emphasised the fact that the models needed to tell the users when to trust their outputs. For this reason, we decided to keep them, even though there are no proven ways to compute them yet.

\subsubsection*{Explanation Aspects.}
The Explanation aspects component was added to the original framework (see \Cref{sec:method:conceptual_components}). This component groups the properties that measure the quality of the generated explanation. These concepts have been generally associated with Functionality-Grounded evaluation because these properties can be measured with metrics at the abstraction level, that is, without the need for users.  

From the analysis, eight properties were found. Necessity, Sufficiency and Contrastivity specifically measure the quality of the selected causes. Their goal is to evaluate whether the reasons the XAI method is providing clearly inform the prediction that was made. Correctness and Completeness are analogous to precision and recall in AI performance metrics. Correctness describes whether the XAI method selected the causes that the AI model used to make a prediction. For explanations generated using the AI model parameters, such as linear regression, the correctness will always be high. Explanations generated by surrogate models will have lower correctness. Completeness quantifies if all the causes that the model used to generate the prediction are present in the explanation.
Representativeness determines whether the explanations are unique to each instance or they generalise over multiple instances. This property helps to estimate the Cognitive Load the users will face when using the system. Size and Structure evaluate the explanations' length and organisation, which affects how easy it will be for users to understand them. 

\subsubsection*{Subjective System Aspects.}
Subjective System Aspects (SSA) are ``users' perceptions of the Objective System Aspects'' \cite{Knijnenburg2015EvaluatingExperiments}. These properties provide evidence that the users perceive the Objective System Aspects. In this modified framework, they help to establish whether the users perceive the OSAs and the Explanation Aspects. Additionally, this component helps us to understand the pertinence of the generated explanations to the users' situational context. These properties are mostly measured at the communication level, but some of them have measures at the abstraction and format level that can be used as proxies of the real value. 

\begin{figure}
\centering
\includegraphics[width=0.7\linewidth]{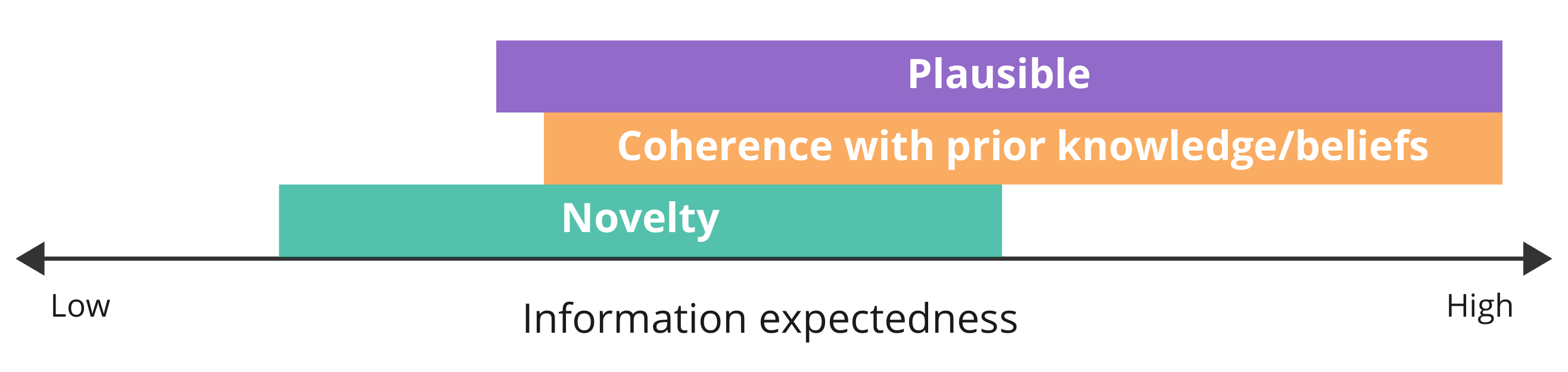}
  \caption{Relation between plausibility, coherence with prior knowledge/beliefs and Novelty with Information Expectedness. Each bar represents the amount of new information that according to the concepts relates to user acceptance.}
\label{fig:information_expectedness}
\vspace{-0.5cm}
\end{figure}

The analysis yielded seven properties for this component. Explanation power measures the perceived quality of the selected causes. Explanations with high power provide valuable justifications for the AI model behaviour. Form of cognitive chunks estimates the semantics of the information provided by the explanation. This concept was coined by \citet{Doshi-Velez2017TowardsLearning} and it has been widely used in the XAI domain. 
Information expectedness measures whether the explanation provides new knowledge to the user. The analysed works used three concepts for this notion: plausibility, coherence with prior knowledge/beliefs, and novelty. 
We decided to keep these notions under one umbrella term because 
we found that they are part of the same scale (see \Cref{fig:information_expectedness}). The relation of each concept with information expectedness is the following:
\begin{itemize}
\item Plausibility \cite{Moreira2022BenchmarkingBox, Beckh2022ALearning, Carvalho2019MachineMetrics}: if the information is expected, the user will think it is plausible. However, the contrary does not necessarily holds. The information can be new but still plausible in the user's mind. 

\item Coherence with prior knowledge/beliefs \cite{Carvalho2019MachineMetrics, Nauta2022FromAI, Sokol2020ExplainabilityApproaches}: the information provided by the explanation should have some level of connection to the user's background. If that relation does not exist, it will be hard for the user to understand the explanation.

\item Novelty \cite{Sokol2020ExplainabilityApproaches, Liao2022ConnectingAI, Lofstrom2022AMethods, Carvalho2019MachineMetrics, Miller2019}: explanations should focus on abnormal causes \cite{Miller2019} and provide information the user does not expect to increase her engagement with the system. However, if the reasons are too unexpected, the user will probably dismiss them and ignore the system.
\end{itemize}

Perceived model competence evaluates whether the user thinks the AI model can perform as expected. The Cognitive load measures the cognitive effort the user makes to understand the explanations.

The last two properties measure the fit between the explanation and the situational context. Relevance to the task measures whether the explanation provides insights that help to perform the task better. An explanation has to be relevant to be useful for the task the user has to perform; otherwise, she will not exploit it. For example, in a medical context, this would measure whether the explanations are actionable in the patient's state. Alignment with situational context evaluates whether the provided explanation is appropriate for the usage context. For instance, a complex visualisation cannot be used correctly in a time-constrained context.  
{\footnotesize

\begin{xltabular}{\textwidth}{@{}p{2.5cm}Xp{2cm}@{}}
\caption{Table of all explanation properties and their definitions based on the reviewed literature.} \label{tab:properties} \\

\toprule 
\textbf{Property} & \textbf{Definition} & \textbf{References}\\ \hline 
\endfirsthead

\multicolumn{3}{c}%
{\tablename\ \thetable{} -- continued from previous page} \\

\hline 
\textbf{Property} & \textbf{Definition} & \textbf{References}\\ \hline 
\endhead

\hline 

\endfoot

\hline
\endlastfoot

\multicolumn{3}{@{}l}{\highlight[objective]{Objective system aspects}}\\[5pt]
  AI Model performance
  & The accomplishment level the AI model has with respect to the task for which it was trained.  
  & \\
  
  AI Model certainty
  & The confidence the AI model has in its prediction.  
  & \cite{Liao2022ConnectingAI,Carvalho2019MachineMetrics,Nauta2022FromAI, Vilone2021NotionsIntelligence} \\

  Certainty  
  & The confidence the XAI method has in the explanation.  
  & \cite{Hoffman2018MetricsProspects, Carvalho2019MachineMetrics, Nauta2022FromAI, Liao2022ConnectingAI, Beckh2022ALearning, Tintarev2015ExplainingEvaluation}\\
  
  Continuity 
  & The function should provide similar explanations for similar instances. 
  & \cite{Hsiao2021RoadmapXAI, Coroama2022EvaluationXAI, Hoffman2018MetricsProspects, Carvalho2019MachineMetrics, Markus2020TheStrategies} \\
  
  Separability & The XAI method should return different explanations for different instances. & \cite{Carvalho2019MachineMetrics}\\
  
  Consistency 
  & The degree to which different runs of the XAI method yield similar XAI functions.
  & \cite{Hsiao2021RoadmapXAI, Carvalho2019MachineMetrics, Nauta2022FromAI, Tonekaboni2019WhatUse, Liao2022ConnectingAI, Vilone2021NotionsIntelligence, Beckh2022ALearning, Lofstrom2022AMethods}\\

\midrule
\multicolumn{3}{@{}l}{\highlight[explanation]{Explanation aspects}} \\[5pt]
  Necessity & Measures whether the explanation method selected the causes that are responsible for the prediction. If the necessary causes change, then the prediction will also change.  & \cite{Miller2019, Liao2022ConnectingAI, Nauta2022FromAI, Carvalho2019MachineMetrics} \\

  Sufficiency & 
  Measures whether the explanation method did not select causes that do not affect the prediction. If non-selected causes change, the prediction would still hold, and thus the explanation should not change. & \cite{Miller2019} \\

  Correctness &  Quantifies the extent to which the selected causes are correct with respect to the model reasoning
   & \cite{Liao2022ConnectingAI, Carvalho2019MachineMetrics, Nauta2022FromAI, Markus2020TheStrategies, Vilone2021NotionsIntelligence, Sokol2020ExplainabilityApproaches} \\

  Completeness & Quantifies if all the causes that the model used to generate the prediction are present in the explanation. & \cite{Carvalho2019MachineMetrics, Nauta2022FromAI, Liao2022ConnectingAI, Markus2020TheStrategies, Vilone2021NotionsIntelligence} \\

  Contrastivity & Measures whether the explanation contains reasons that highlight differences with respect to other possible outcomes. Low contrastivity will provide the same reasons for instances in which the model predicts different classes. & \cite{Carvalho2019MachineMetrics, Nauta2022FromAI, Miller2019} \\

  Size & Refers to the amount of information present in the explanation. & \cite{Nauta2022FromAI, Carvalho2019MachineMetrics, Vilone2021NotionsIntelligence, Markus2020TheStrategies, Liao2022ConnectingAI, Doshi-Velez2017TowardsLearning} \\
  
  Structure & The information should be displayed in a way that allows the users to understand the hierarchy of the information quickly & \cite{Nauta2022FromAI, Hsiao2021RoadmapXAI, Carvalho2019MachineMetrics, Tonekaboni2019WhatUse, Tintarev2015ExplainingEvaluation, Markus2020TheStrategies, Vilone2021NotionsIntelligence, Sokol2020ExplainabilityApproaches} \\

  Representativeness & An explanation is representative if it holds for many distinct but similar instances. & \cite{Carvalho2019MachineMetrics, Coroama2022EvaluationXAI, Vilone2021NotionsIntelligence, Sokol2020ExplainabilityApproaches} \\
   
\midrule
\multicolumn{3}{@{}l}{\highlight[subjective]{Subjective system aspects}}\\[5pt]
    Explanation power & Measures whether the selected causes make the user understand the reasons the model considered when making a decision & \cite{Liao2022ConnectingAI, Vilone2021NotionsIntelligence, Carvalho2019MachineMetrics, Wanner2022ALearning} \\

    Form of cognitive chunks & Refers to the semantics and structure of the pieces of information the user will receive. & \cite{Doshi-Velez2017TowardsLearning, Nauta2022FromAI, Vilone2021NotionsIntelligence, Carvalho2019MachineMetrics}\\

    Information expectedness & Level of surprise of the information revealed by the explanation & \cite{Hsiao2021RoadmapXAI, Carvalho2019MachineMetrics,Nauta2022FromAI, Tonekaboni2019WhatUse, Miller2019, Liao2022ConnectingAI, Vilone2021NotionsIntelligence, Sokol2020ExplainabilityApproaches, Beckh2022ALearning, MoraffahCausalEvaluation,Lofstrom2022AMethods, Moreira2022BenchmarkingBox}\\
    
    Perceived model competence & Measures the user's impression of the model competence for the task at hand & \cite{Chen2005TrustAgents} \\

    Cognitive Load & Refers to the cognitive effort the user has to do to achieve the task. & \cite{Carvalho2019MachineMetrics, Liao2022ConnectingAI, Vilone2021NotionsIntelligence}\\
   
    Relevance to the task & Level of explanation usefulness to the user's task. & \cite{Hsiao2021RoadmapXAI, Nauta2022FromAI, Tonekaboni2019WhatUse, Miller2019, Liao2022ConnectingAI, Vilone2021NotionsIntelligence, Moreira2022BenchmarkingBox, Wanner2022ALearning}  \\
    
    Alignment with situational context & Level of appropriateness of the explanation to the usage context& \cite{Hsiao2021RoadmapXAI, Carvalho2019MachineMetrics, Tonekaboni2019WhatUse, Liao2022ConnectingAI, Sokol2020ExplainabilityApproaches}\\

\midrule
\multicolumn{3}{@{}l}{\highlight[experience]{User experience}}\\[5pt]
     Curiosity &  Measures whether the user is intrinsically motivated to understand the explanation. If the user is curious, she will be more attentive to the task and, therefore, more engaged with the system. & \cite{Hoffman2018MetricsProspects, Vilone2021NotionsIntelligence, Hsiao2021RoadmapXAI} \\

    Satisfaction & Refers to the level of fulfilment the user gets while interacting with the system. This satisfaction is always measured at the communication level because it is for the users to decide whether they feel good about the overall system interaction. & \cite{Hsiao2021RoadmapXAI, Hoffman2018MetricsProspects, Tintarev2015ExplainingEvaluation, Vilone2021NotionsIntelligence, Wanner2022ALearning}\\
  
    Trust & We use the definition by \citet{Tintarev2015ExplainingEvaluation}: ``perceived confidence in a system's competence'' &  \cite{Hsiao2021RoadmapXAI, Hoffman2018MetricsProspects, Tonekaboni2019WhatUse, Liao2022ConnectingAI, Tintarev2015ExplainingEvaluation, Chen2005TrustAgents, Ashoori2019InProcesses, Wanner2022ALearning} \\

    Understanding & Refers to the ability of the user to interpret the system's output correctly. The user fails to understand when she cannot interpret or incorrectly interprets the system's explanation and prediction. This involves the creation of the user's mental model and how that aligns with the system's functionality. & \cite{Carvalho2019MachineMetrics,  Hoffman2018MetricsProspects, Hsiao2021RoadmapXAI, Lofstrom2022AMethods, Markus2020TheStrategies, Vilone2021NotionsIntelligence, Wanner2022ALearning}\\

    Usefulness & Measures whether the explanation helps the user to understand the AI prediction.  & \cite{Beckh2022ALearning, Tintarev2015ExplainingEvaluation, Vilone2021NotionsIntelligence, Wanner2022ALearning} \\

    Controllability & Measures whether the user perceives she has some level of control over the system. This could manifest as the ability to reverse actions, correct the system, filter or zoom the explanation, or ask questions to clarify the explanation or prediction. & \cite{Hoffman2018MetricsProspects, Hsiao2021RoadmapXAI, Liao2022ConnectingAI, Nauta2022FromAI, Sokol2020ExplainabilityApproaches, Tintarev2015ExplainingEvaluation, Vilone2021NotionsIntelligence}\\
\midrule
\multicolumn{3}{@{}l}{\highlight[interaction]{Interaction}}\\[5pt]
   Efficiency & Measures the speed at which a task can be performed. & \cite{Vilone2021NotionsIntelligence, Tintarev2015ExplainingEvaluation}\\
    
    Performance & Measures how well the user can do the task while using the system (prediction+explanations). & \cite{Lofstrom2022AMethods, Hsiao2021RoadmapXAI, Hoffman2018MetricsProspects} \\
    
    Reliance & Measures whether the user is willing to provide control to the machine for the given task. & \cite{Hoffman2018MetricsProspects, Hsiao2021RoadmapXAI, Tintarev2015ExplainingEvaluation, Vilone2021NotionsIntelligence} \\

\bottomrule
\end{xltabular}

}

\subsubsection*{User experience.}
The User experience factors evaluate what the user encounters when interacting with the system \cite{Knijnenburg2015EvaluatingExperiments}. The analysis did not find surprising aspects because all but Curiosity, have been studied in recommender systems. The aspects found are Satisfaction, Trust, Understanding, Usefulness, Controllability and Curiosity. This last one was highlighted as extremely relevant in the case of explanations because the search for an explanation is modulated by the user's curiosity \cite{Hsiao2021RoadmapXAI}. Moreover, the motivation to ask or explore an explanation is determined by the user's curiosity \cite{Hoffman2018MetricsProspects}.

\subsubsection*{Interaction.}
Interaction factors measure aspects related to the possible adoption of the system. Three properties were found to be relevant: Efficiency, Performance and Reliance.
Efficiency measures how fast the user can perform the task. Performance evaluates the level of achievement the user reaches while using the system. Finally, Reliance measures to which extent the user is willing to provide control to the AI model to perform the task. 

\subsection{Relations Between Properties}\label{sec:framework_relations}

\begin{figure}[h]
\centering
\includegraphics[trim={0 1cm 0 0},clip, width=0.8\textwidth ]{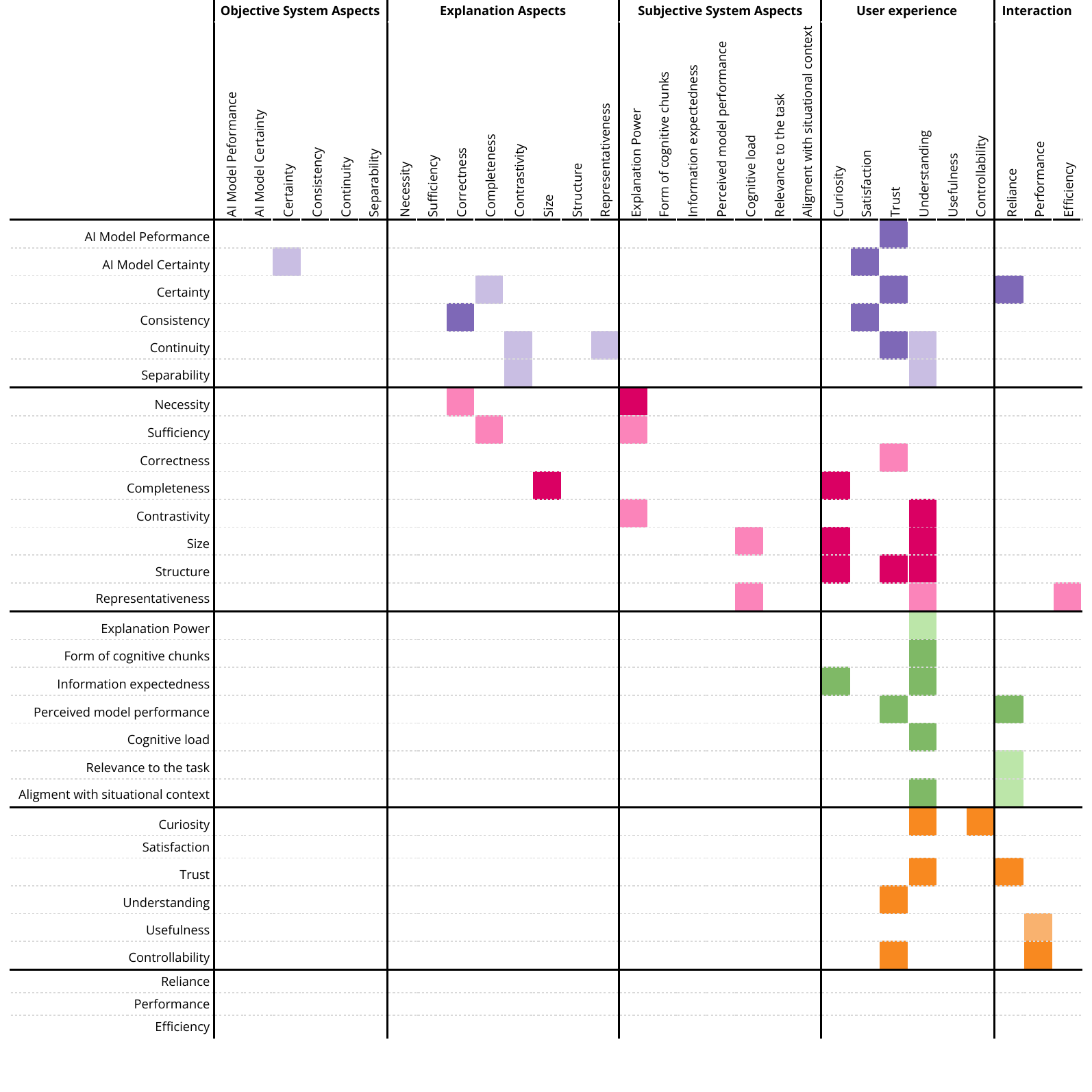}

  \caption{Relations between properties. The relations are directed: horizontal properties are the source, and vertical properties receive the effect. High saturation squares indicate that the relationship has been described in the literature, and low saturation squares indicate the relation was inferred from the definitions.}
\label{fig:relations_summary}
\vspace{-.5cm}
\end{figure}

Explanation properties are related to each other in intricate ways \cite{Knijnenburg2015EvaluatingExperiments}. As stated in \Cref{sec:method:conceptual_components}, by scanning past research, we identified such relationships and linked the properties in our framework as described in the literature; for instance, explanation size affects user curiosity \cite{Hoffman2018MetricsProspects}. 
In this way, we mapped out relations proposed in the literature and relations inferred from the properties' definitions. \Cref{tab:relations} describes the relations found for each property and \Cref{fig:relations_summary} displays a visual summary of the interactions. These relations help theorise the expected causal effects between the properties. In practical terms, they serve as hypotheses for the Structural Equation Model.

{\footnotesize

\begin{xltabular}{\textwidth}{@{}p{2.8cm}X@{}}
\caption{Relations between properties. Relations without reference were hypothesised based on the properties' definitions.} \label{tab:relations} \\

\toprule
\textbf{Explanation Property} & \textbf{Relations with other properties}\\ \hline 
\endfirsthead

\multicolumn{2}{c}%
{\tablename\ \thetable{} -- continued from previous page} \\

\hline \textbf{Explanation Property} & \textbf{Relations with other properties}\\ \hline 
\endhead

\hline 

\endfoot

\hline
\endlastfoot
  
\multicolumn{2}{@{}l}{\highlight[objective]{Objective system aspects}}\\[5pt]
Model performance & The performance of the AI will affect the level of Trust the users can achieve \cite{Tintarev2015ExplainingEvaluation, Dominguez2019TheImages} \\

Model certainty  & If the AI model is uncertain of the predictions, the XAI method will have more difficulties obtaining consistent explanations, which will affect the XAI method's certainty. This confidence will also affect the satisfaction with the system because, as \citet{Tintarev2015ExplainingEvaluation} explains, a user might be more forgiving if the system admits it is not confident about a prediction.\\

Certainty & If the explanation shows its limitations, the user may not relay or trust the system \cite{Hoffman2018MetricsProspects, Liao2022ConnectingAI}. Low certainty will affect the correctness of the explanation.\\

Continuity & Higher continuity increases the understanding of the model because the similarity of explanations helps to learn from the model. It also helps to produce contrastive and representative explanations. Ultimately, high continuity can also increase Trust \cite{Tonekaboni2019WhatUse}.\\

Separability & Higher separability increases the understanding of the model and the contrastivity of the explanations. \\

Consistency & Low consistency may decrease user satisfaction \cite{Hsiao2021RoadmapXAI} and correctness \cite{Liao2022ConnectingAI}\\

\midrule
\multicolumn{2}{@{}l}{\highlight[explanation]{Explanation Aspects}}\\[5pt]

Necessity & Affects the explanation power \cite{Liao2022ConnectingAI}. Additionally, if the necessary causes are selected, then the correctness will be high.\\

Sufficiency & Affects the explanation power. Moreover, if the sufficient causes are selected the completeness of the explanation will increase. \\

Correctness & An explanation with high correctness will faithfully reflect the decision process of the AI model. This could increase Trust in the explanation and AI model \\

Completeness & The explanation size is related to completeness: the bigger the explanation, the more complete it will be \cite{Liao2022ConnectingAI, Carvalho2019MachineMetrics}. However, bigger explanations might decrease curiosity \cite{Hsiao2021RoadmapXAI}. \\

Contrastivity & High contrastivity will increase the explanation power. Additionally, this property will affect understanding because the people expect explanations to be contrastive \cite{Miller2019}. \\

Size & The amount of information affects curiosity in an inverted U-shaped pattern: little or excessive information reduces curiosity \cite{Hsiao2021RoadmapXAI}. The size of the explanation also affects how easily a user can understand the explanation \cite{Liao2022ConnectingAI, Markus2020TheStrategies}. This last effect could be mediated by the Cognitive Load.\\

Structure & The design of the information that will be shown affects its trustworthiness \cite{Tintarev2015ExplainingEvaluation}, curiosity \cite{Hsiao2021RoadmapXAI} and ultimately how easy they can be understood \cite{Markus2020TheStrategies, Sokol2020ExplainabilityApproaches}.  \\

Representativeness & This property affects understanding, cognitive load and efficiency because the user can understand an explanation more quickly if it is similar to those she has seen before. \\

\midrule
\multicolumn{2}{@{}l}{\highlight[subjective]{Subjective System Aspects}}\\[5pt]

Explanation power & The quality of the selected causes will help increase understanding. \\

Form of cognitive chunks &  It affects understanding because this property measures how interpretable are the information pieces the user receives \cite{Carvalho2019MachineMetrics} \\

Information expectedness & If the information is coherent with the user's beliefs, they will be more likely to understand it \cite{Sokol2020ExplainabilityApproaches}. However, if the information does not add anything new to their existing knowledge, they are less likely to be curious \cite{Sokol2020ExplainabilityApproaches}.   \\

Perceived model competence & When the users perceive the AI model can perform, they are more likely to trust it and eventually to rely on it \cite{Chen2005TrustAgents}.\\

Cognitive Load & An explanation with low cognitive load will be easier to understand \cite{Liao2022ConnectingAI}.\\

Relevance to the task & If the explanations help the development of the task, the user is more likely to rely on the AI advice. \\

Alignment with situational context & Trust in the system is context-dependent. If the system is aligned with the situation the user has to perform, she will be more likely to trust it. \cite{Hoffman2018MetricsProspects}. Additionally, this could build up until the user starts to rely on the AI agent.\\

\midrule
\multicolumn{2}{@{}l}{\highlight[experience]{User experience}}\\[5pt]

Curiosity & Mental model formation, which is the final goal of understanding, is modulated by Curiosity \cite{Hsiao2021RoadmapXAI}. Additionally, Curiosity encourages users to explore and interact with the system \cite{Hsiao2021RoadmapXAI}.  \\

Satisfaction & \\

Trust & Reliance is an outcome of appropriate trust \cite{Tintarev2015ExplainingEvaluation, Hsiao2021RoadmapXAI, Chen2005TrustAgents, Hoffman2018MetricsProspects}. Mental model formation is also modulated by Trust in the system \cite{Hsiao2021RoadmapXAI}\\

Understanding & If users cannot understand the behaviour, Trust will be lost \cite{Hsiao2021RoadmapXAI}\\

Usefulness & If the user finds the explanations helpful, they are more likely to increase the user performance with the system.\\

Controllability & The possibility of interaction increases Trust in the system \cite{Tintarev2015ExplainingEvaluation}. Good interaction with the system can increase the performance of the users \cite{Hsiao2021RoadmapXAI}\\

\bottomrule
\end{xltabular}

}

\subsection{Measurements}\label{sec:framework_measurements}
\begin{figure}[t]
\centering
\includegraphics[trim={0 0.5cm 0 0},clip,width=\textwidth ]{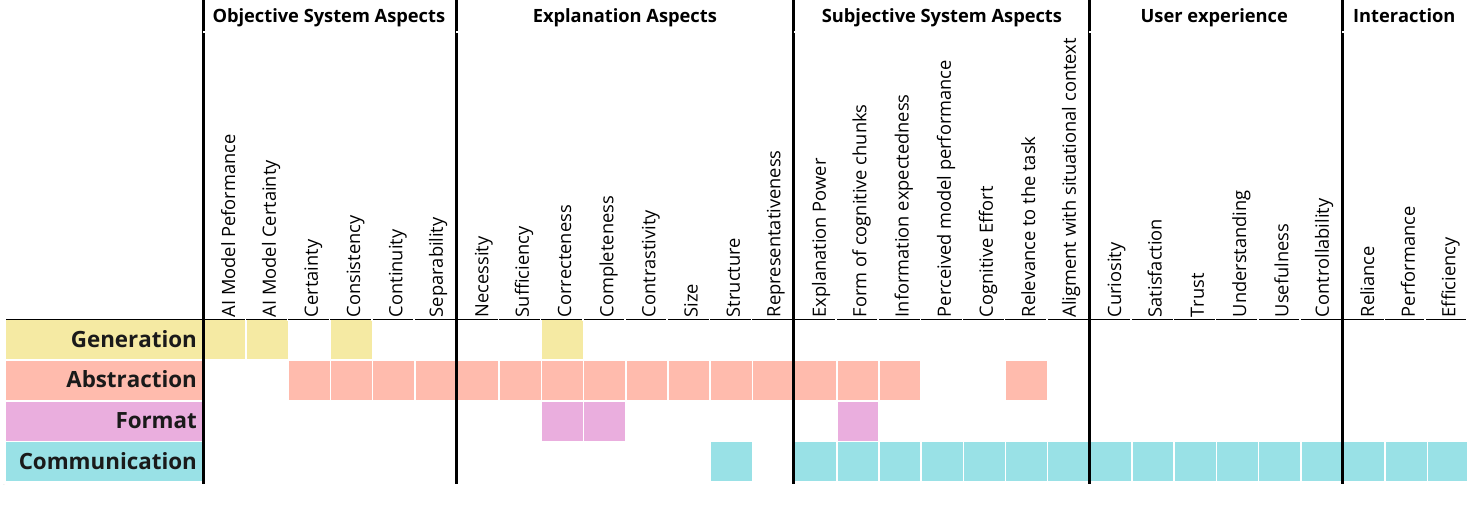}
  \vspace{-0.5cm}
  \caption{Existence of a measurement procedure for each property in each explanation element. Each coloured rectangle indicates that a measurement has been defined for the tuple (property, explanation element) }
\label{fig:metrics_summary}
\vspace{-0.5cm}
\end{figure}
As explained in \Cref{sec:method:explanation_elements}, we classified measurements of properties with three criteria: property they measure, explanation element in which they are applied and type of procedure. 
The explanations elements allow us to capture the complexity of the explanations: they are not simply an object we show to users; a model has generated them and then transformed them to be shown to users in a specific situational context. The four elements are:

\begin{itemize}

\item \textbf{Generation element}. Refers to the process that was conducted to select the causes that will be displayed in the explanation for a specific object. The measures of this element are applied to the XAI function and AI model. They check the function's parameters to obtain indicators. 

\item \textbf{Abstraction element}. Represents the selected causes of the explanation without considering the format in which they will be displayed. For example, for feature importance, this could be a table with the features and their corresponding importance values. Measurements that are applied at this level look at the data that was selected by the XAI method as an explanation.

\item \textbf{Format element}. Refers to the manner in which the causes will be presented to the user. This could be as example-based, text, visualisation, etc. In this study, few measurements were found to be applicable at this level. However, each specific media type has its own measurements that could be modified to be applied. For instance, for visual explanations, the data-ink ratio could be applied to analyse whether the most important features use more ink in the visualisation.

\item \textbf{Communication element}. It refers to the process of interacting with the formatted explanation. During this process, information can be captured as interaction measurements, as well as self-reported information.
\end{itemize}

In \Cref{fig:metrics_summary}, a coloured square is present if at least one measurement exists for that (property, element) tuple. It is noted that the User Experience and Interaction conceptual components are only measured at the communication level, i.e. only when the explanation is displayed to users. What stands out in this figure is the number of measurements at the abstraction level for the Subjective System Aspects component. In Knijnenburg's framework \cite{Knijnenburg2015EvaluatingExperiments}, these aspects were recommended to be measured with self-reporting questionnaires. However, our analysis found that for some of them, metrics have been proposed at the Abstraction element, which means that some computational metric is applied to the abstract explanation to obtain a value. 

The main advantage of decoupling the properties from the ways to measure them is that it allows researchers to select measurements considering the study constraints. For instance, if the study is conducted with users that do not have much time to answer questionnaires, and the researchers want to measure \highlight[subjective]{Explanation power}, \highlight[subjective]{Form of cognitive chunks}, \highlight[experience]{Curiosity} and  \highlight[experience]{Understanding}, they may choose to measure the first two properties at the abstract level of the explanation and the last two at the communication level with questionnaires. In this way, they do not overwhelm the users with questions but still measure the required properties. 

As pointed out before, we also classified the measurements by the type of procedure. In this analysis, only procedures that produce a quantitative value were considered. This means that qualitative interviews were not considered, nor were experiment tasks. The four types are:

\begin{itemize}
\item \textbf{Quantitative interviews}: closed-ended questions, usually in Likert scale.
\item \textbf{Computational metrics}: mathematical functions that are applied to the explanations or XAI methods.
\item \textbf{Behaviour metrics}: indicators of user behaviour and interaction with a system. For example, the number of interactions within the system and the time to complete a task.
\item \textbf{Objective Body Measurements}: measurements taken from the user body. The most common is eye-tracking.
\end{itemize}

{\footnotesize
\begin{xltabular}{\textwidth}{@{}p{3.8cm}X@{}}
\caption{Measurements of properties} \label{tab:metrics} \\

\toprule
\textbf{Explanation Property} & \textbf{Measurement}\\ \hline 
\endfirsthead

\multicolumn{2}{c}%
{\tablename\ \thetable{} -- continued from previous page} \\

\hline \textbf{Explanation Property} & \textbf{Measurement}\\ \hline 
\endhead

\hline 

\endfoot

\hline
\endlastfoot

\multicolumn{2}{@{}l}{\vspace{0.1ex}\colorbox{vanillaf}{\textit{Generation}}}\\[5pt]
  AI Model performance & Measured by the model type appropriate metrics: accuracy, f-score, precision, recall and others. \\
  
 AI Model certainty &  This property can be measured for each individual prediction and the global model. If it is measured globally, it should be measured over a dataset similar to the data the real system will face. \cite{Coroama2022EvaluationXAI} \\

  Consistency & Implementation invariance: check whether the XAI function parameters are the same after different runs of the XAI method creation \cite{Carvalho2019MachineMetrics, Vilone2021NotionsIntelligence}  \\
 
    Correctness &  Translucency \cite{Carvalho2019MachineMetrics}\\
\midrule
\multicolumn{2}{@{}l}{\vspace{0.1ex}\colorbox{melonf}{\textit{Abstraction}}}\\[5pt]

  Certainty & Confidence Accuracy \cite{Nauta2022FromAI}  \\
   
  Continuity &  Connectedness \cite{Nauta2022FromAI} also in \cite{Singh2022ExplainabilityMethods, MoraffahCausalEvaluation, Moreira2022BenchmarkingBox, Vilone2021NotionsIntelligence, Ge2021CounterfactualAI, Carlevaro2022CounterfactualDescription}; Stability for Slight Variations \cite{Nauta2022FromAI}; Fidelity for Slight Variations \cite{Nauta2022FromAI}  \\

  Separability & Separability \cite{Carvalho2019MachineMetrics} \\

  Consistency & Stability of explanation: check whether the explanations for a single object change for different instances of the XAI method \cite{Coroama2022EvaluationXAI, Carvalho2019MachineMetrics} \\
  Necessity & Responsability of an outcome \cite{Miller2019}; Sparsity and Sparsity rate \cite{Moreira2022BenchmarkingBox}; Deletion Check \cite{Nauta2022FromAI, Li2022AMetrics} \\

  Sufficiency & Count whether the AI model prediction changes when the non-selected causes change \cite{Miller2019} \\
  
  Correctness &  Model Parameter Randomization Check, Explanation Randomization, White Box Check, Controlled Synthetic Data Check, Predictive Performance \cite{Nauta2022FromAI}; Fidelity  \cite{Sokol2020ExplainabilityApproaches, Coroama2022EvaluationXAI, Nauta2022FromAI}; Alignment between AI model features and explanation features \cite{Li2022AMetrics, Velmurugan2021DevelopingLearning}\\

  Completeness & Preservation Check \cite{Nauta2022FromAI}; Completeness \cite{Coroama2022EvaluationXAI, MoraffahCausalEvaluation}; Recall \cite{Velmurugan2021DevelopingLearning}  \\

  Contrastivity & Data Randomization, Target Sensitivity, Target Dicriminativeness \cite{Nauta2022FromAI}; Sensitivity \cite{Vilone2021NotionsIntelligence} \\

  Size & Total size or sparsity \cite{Nauta2022FromAI} \\
  
  Structure & Incremental Deletion \cite{Nauta2022FromAI, Coroama2022EvaluationXAI}; Covariate Regularity \cite{Nauta2022FromAI, Coroama2022EvaluationXAI}; Chronology \cite{Sokol2020ExplainabilityApproaches}; Single Deletion \cite{Nauta2022FromAI} \\

  Representativeness & Explanation support (number of instances to which the explanation applies over the number of instances) \cite{Sokol2020ExplainabilityApproaches, Carvalho2019MachineMetrics, Vilone2021NotionsIntelligence, Coroama2022EvaluationXAI} \\
  

    Explanation power & Sensitivity Axiom \cite{Carvalho2019MachineMetrics}\\

    Form of cognitive chunks & Covariate Homogeneity \cite{Nauta2022FromAI} \\

    Information expectedness & Alignment with Domain Knowledge \cite{Nauta2022FromAI} \\

    Relevance to the task & Pragmatism \cite{Nauta2022FromAI, Coroama2022EvaluationXAI, Moreira2022BenchmarkingBox}; Attribute costs \cite{Vilone2021NotionsIntelligence} \\

\midrule
\multicolumn{2}{@{}l}{\vspace{.1ex}\colorbox{plumf}{\textit{Format}}}\\[5pt]
Correctness &  Percentage of invalid rules \cite{Vilone2021NotionsIntelligence} \\

  Completeness &  Rules redundancy \cite{Vilone2021NotionsIntelligence} \\
 Form of cognitive chunks & BLEU and METEOR \cite{Vilone2021NotionsIntelligence}; Perceptual Realism \cite{Nauta2022FromAI}  \\

     \midrule

 \multicolumn{2}{@{}l}{\vspace{.1ex}\colorbox{skybluef}{\textit{Communication}}}\\[5pt]

  Structure & Questionnaire \cite{Vilone2021NotionsIntelligence, Chen2005TrustAgents} \\
  Explanation power &  Questionnaire \cite{Vilone2022AIntelligence}\\

    Form of cognitive chunks &  Perceived Homogeneity \cite{Nauta2022FromAI} \\

    Information expectedness &  Questionnaire \cite{Vilone2022AIntelligence}\\
    
    Perceived model competence &  Questionnaire \cite{Ashoori2019InProcesses}\\

    Cognitive Load &  NASA TLX \cite{Hsiao2021RoadmapXAI}\\
   
    Relevance to the task &  Questionnaire \cite{Vilone2022AIntelligence}\\
    
    Alignment with situational context & Goodness explanation \cite{Hoffman2018MetricsProspects} \\
    Curiosity & Curiosity Checklist \cite{Hoffman2018MetricsProspects}; Eye Movement Pattern \cite{Hsiao2021RoadmapXAI} \\

    Satisfaction & Explanation Satisfaction Scale \cite{Hoffman2018MetricsProspects}; Eye Movement Pattern \cite{Hsiao2021RoadmapXAI}; Loyalty \cite{Tintarev2015ExplainingEvaluation}; Questionnaire \cite{Ashoori2019InProcesses} \\
  
    Trust & Trust Scale \cite{Hoffman2018MetricsProspects}; Questionnaire \cite{Ashoori2019InProcesses, Chen2005TrustAgents, Pu2007Trust-inspiringSystems, Vilone2022AIntelligence}\\

    Understanding & Questionnaires \cite{Tintarev2015ExplainingEvaluation,Ashoori2019InProcesses, Vilone2022AIntelligence}\\

    Helpfulness & Questionnaires \cite{Tintarev2015ExplainingEvaluation, Vilone2022AIntelligence}; Evaluate user action before and after explanation \cite{Tintarev2015ExplainingEvaluation}\\

    Controllability & Concept-level feedback Satisfaction Ratio \cite{Chen2020TowardsRecommendation}; The extent to which a user can produce certain outcomes \cite{Hoffman2018MetricsProspects}\\
    Efficiency & Interaction time and number of interactions to perform a task \cite{Tintarev2015ExplainingEvaluation}\\
    
    Performance & Performance metrics with respect to the primary goal \cite{Hoffman2018MetricsProspects} \\
    
    Reliance & Questionnaires \cite{Ashoori2019InProcesses, Chen2005TrustAgents}; Willingness to accept AI agent advice \cite{Hsiao2021RoadmapXAI} \\
  \bottomrule
\end{xltabular}

}

The metrics for each (property, element) are listed in \Cref{tab:metrics}. This table was built under the following rules:
\begin{itemize}
\item Several measurements have been defined in multiple works. To avoid naming all of them in the tables, we built upon existing work by using the name proposed by \citet{Nauta2022FromAI} to summarise metrics every time a similar metric was defined in another work. This new work was added as a reference under the same name.
\item Some procedures were not described with a specific name in the paper. In those cases, an explanatory sentence was used to name them. 
\item If asking questions was proposed as a procedure, but no measurement model or questions were provided, the measurement was not considered.
\item For a given property, questions proposed in different papers were joined together under the \textit{Questionnaire} term.
\end{itemize}










\section{Illustrative Example}
In this section, we provide an illustrative example of how our framework can be used. Researchers have an AI model that predicts whether a patient will be readmitted to the emergency department within 30 days. \colorbox{vanillaf}{\textit{SHAP}} \cite{Lundberg2018ExplainableSurgery} is used to determine the \colorbox{melonf}{\textit{feature importance}} on a patient level and this information is then visualised in a \colorbox{plumf}{\textit{force plot}} \cite{Lundberg2018ExplainableSurgery}. Finally, medical staff \colorbox{skybluef}{\textit{analyses}} the prediction and visual explanation to decide whether they discharge a patient.

In this context, assessing the explanation requires several steps. First, researchers have to decide which properties to measure. This is a decision support system, so according to \cite{Liao2022ConnectingAI}, the most relevant properties would be \highlight[experience]{\textit{Trust}}, \highlight[experience]{\textit{Controllability}}, and \highlight[experience]{\textit{Understanding}}. The researchers conjecture that \highlight[interaction]{\textit{Reliance}} and \highlight[interaction]{\textit{Performance}} will be good indicators of adoption. 
Second, they have to select explanation properties that relate to these five properties. Following the theoretical causal relations in \Cref{tab:relations} and \Cref{fig:relations_summary}, such properties are: 
\begin{itemize}
\item \highlight[objective]{\textit{AI Model performance}},  \highlight[objective]{\textit{Certainty}}, \highlight[objective]{\textit{Continuity}}
\item \highlight[explanation]{\textit{Size}}, \highlight[explanation]{\textit{Structure}}, \highlight[explanation]{\textit{Representativeness}}
\item \highlight[subjective]{\textit{Form of cognitive chunks}}, \highlight[subjective]{\textit{Information expecteness}},  \highlight[subjective]{\textit{Cogntive load}}, \\\highlight[subjective]{\textit{Perceived model competence}}, \highlight[subjective]{\textit{Aligment with situational context}}
\item \highlight[experience]{\textit{Curiosity}}
\end{itemize}

Finally, to assess all selected properties, researchers pick appropriate metrics from  \Cref{tab:metrics}. The metrics' scores applied to the elements \colorbox{melonf}{\textit{abstraction}} and \colorbox{plumf}{\textit{format}} are averaged over the single explanations, and the questionnaires are applied at the end of the experience. This data is then analysed using structural equation modelling.

\section{Conclusion and Future Work}

In this work, we have presented a user-centric evaluation framework for XAI inspired by research on recommender systems allowing researchers to conduct systematic user experience evaluations in the context of XAI-based systems. Our proposal integrates the current state of the art in XAI evaluation but also allows to easily incorporate new properties or metrics that might become relevant for new applications. By decoupling the aspects of explanations and the procedures to measure them, this framework provides researchers with more tools to choose what and how to measure, and why it is necessary to do it, with the ultimate goal of evaluating the user experience under these new XAI scenarios.

For future work, we plan to validate the framework with user studies. 
We aim at validating metrics, properties, as well as mediation and causal effects between them. Additionally, 
we could include experimental designs that compare different explanations, for instance, by comparing the user experience under two different visualisations for explanations generated with SHAP. Furthermore, we did not analyse how specific situational and personal characteristics affect the properties. This area has been explored \cite{Conati2021TowardSystems, Szymanski2022ExplainingDonts, Lim2009WhySystems}, but more work is needed to connect those findings to explanation properties. Another area of improvement is proposing a standardised report of results to increase fair comparison with previous studies. Lastly, there is no comprehensive survey on the maturity of each of the measurements and on the relations between the properties. Such a survey would help researchers and practitioners to understand the maturity of each property and measurement to help them plan their studies based on current evidence.

 \subsection*{Acknowledgements}

This work was partially funded by ANID Chile, Millennium Science Initiative Program, codes ICN2021\_004 (iHealth) and ICN17\_002 (IMFD), by Basal
Funds for Center of Excellence FB210017 (CENIA), the Research Foundation Flanders (FWO, grant G0A3319N) and KU Leuven  (grant C14/21/072). In addition, we thank Fondecyt grant 1231724.  The research of Ivania Donoso-Guzmán was supported by the doctoral scholarship of ANID Chile.

\newpage
\bibliographystyle{splncs04}

\bibliography{references}

\end{document}